\documentstyle[floats,aps,,psfig,preprint]{revtex}
\begin{document}
\newcommand{\slsh}{\not\!}
\newcommand{\kpp}{k_{++}}
\newcommand{\skpp}{\!\not\!k_{++}}
\newcommand{\kpm}{k_{+-}}
\newcommand{\skpm}{\!\not\!k_{+-}}
\newcommand{\kmp}{k_{-+}}
\newcommand{\skmp}{\!\not\!k_{-+}}
\newcommand{\kop}{k_{0+}}
\newcommand{\kopm}{k_{0+\mu}}
\newcommand{\ga}{\gamma_\alpha}
\newcommand{\gb}{\gamma_\beta}
\newcommand{\gm}{\gamma_\mu}
\newcommand{\svkpp}{\sigma_V(k_{++}^2)}
\newcommand{\sskpp}{\sigma_S(k_{++}^2)}
\newcommand{\svkmp}{\sigma_V(k_{-+}^2)}
\newcommand{\sskmp}{\sigma_S(k_{-+}^2)}
\newcommand{\svkpm}{\sigma_V(k_{+-}^2)}
\newcommand{\sskpm}{\sigma_S(k_{+-}^2)}
\newcommand{\gfv}{\gamma_5}
\newcommand{\ktldm}{\tilde{k}_\mu}
\newcommand{\tr}{{\rm tr}}
\renewcommand{\Im}{\mbox{Im}}
\newcommand{\bea}{\begin{eqnarray*}}
\newcommand{\eea}{\end{eqnarray*}}
\newcommand{\be}{\begin{equation}}
\newcommand{\ee}{\end{equation}}
\newcommand{\g}{\gamma}
\newcommand{\G}{\Gamma}

\title{Off Mass Shell Effects in Hadron Electric Dipole Moments}
\author{M.B.  Hecht, B.H.J.  McKellar}
\date{\today}
\draft
\address{{\it School of Physics, University of Melbourne, 
Parkville VIC 3052,
     Australia}}
\preprint{UM-P-98/56,~RCHEP-98/15}
\maketitle

\begin{abstract}
We note that off the quark mass shell the operators
$(p_i+p_f)_\mu\gamma_5$ and $i\sigma_{\mu\nu}(p_i -
p_f)^\nu\gamma_5$, both of which reduce to $-
\vec{\sigma}\cdot\vec{E}$ in the non-relativistic limit,
are no longer identical.  In this paper we explore the
effects of this difference in the contribution of these
quark electric moments to hadronic electric moments.
\end{abstract}

\newpage
\section{Introduction}
When the loop diagrams which contribute to the electric
dipole moments of quarks are evaluated in theories with $P$
and $T$ violating interactions, in general one finds
a $P$ and $T$ violating quark-photon vertex function
$\tilde{\Gamma}^{\mu_q}$  of the form
\begin{equation}
\tilde{\Gamma}^{\mu_q} = \left[D_{1,q} (p_i+p_f)^\mu\gamma_5
 +  \imath
D_{2,q}\sigma^{\mu\nu}(p_i-p_f)_\nu\gamma_5\right]
 Q_q.\label{new}
\end{equation}
For convenience, we have extracted a factor $Q_q$, the
charge of the quark, in units of the positron charge.
 Heretofore it has been the standard practice to use the
$\gamma_5$ form of the Gordon identity, valid
on mass shell,
\begin{equation}
\bar{u}(p_f)\left[(p_i+p_f)^\mu\gamma_5\right]u(p_i) =
\bar{u}(p_f)\left[
        \imath\sigma^{\mu\nu}q_\nu\gamma_5\right]u(p_i),
\label{gordon}
\end{equation}
with 
\begin{equation}
q^\mu = p_i^\mu - p_f^\mu,
\end{equation}
to write
the quark electric dipole moment interaction in its on mass
shell form 
\begin{equation}
\tilde{\G}^{\mu_{q,oms}} = \imath
D_qQ_q\sigma_{\mu\nu}(p_i-p_f)^\nu\gamma_5, \label{oms}
\end{equation}
and to identify $d_q = D_q Q_q = \left[D_{1,q} + D_{2,q}\right]Q_q$
as the electric dipole moment of the quark.  Note that the
dimension of $d_q$ is $M^{-1}$, and instead of
introducing $d_q$ one could have made the natural units
explicit and introduced a gyroelectric ratio $h_q$,
related to $d_q$ by $d_q = h_q Q_q e /(2m_q)$.  However,
unlike the practice in dealing with magnetic moments, it
is standard practice to use $d_q$ rather than $h_q$.
For further convenience we also write $d_{i,q} =
D_{i,q}Q_q$.

In a previous paper \cite{HechtMcKellar98}, the quark
contribution to the electric and magnetic dipole moments
of the rho meson were calculated using 
vertices and
propagators obtained from studies of the Dyson-Schwinger
and Bethe-Salpeter equations.  This provided some insight
into the effect that QCD and confinement had on the
electromagnetic properties of hadrons.   The electric dipole
moment of the quark used in that calculation was the on
mass shell form of Eq.~(\ref{oms}).

It should be noted that that the T-violating $NN\gamma$ vertex
of equation~(\ref{new}) is gauge invariant only when $p_i^2 = p_f^2$.  However,
this is enough for us to obtain a gauge invariant result
for the $P$ and $T$ violating $\rho - \gamma$ coupling.  As an off shell 
amplitude, it need not be  gauge invariant --- but it must satisfy 
the Ward Takahashi identities.  We will examine the consequences of 
this requirement below.

In this paper we extend that study to examine the effects
of the vertex function ambiguity off shell, and calculate
the electric dipole moment of the rho meson which is
generated by the $P$ and $T$ odd vertex function
$\tilde{\Gamma}^\mu_q$ of Eq.~(\ref{new}).
We continue to use the semi-phenomenological $n$-point
functions of Quantum Chromodynamics obtained in the
Dyson-Schwinger and Bethe- Salpeter framework.  For ease
of comparison with previous results we also give
perturbative, bag model and non-relativistic results, as
in our previous paper \cite{HechtMcKellar98}.  The
methodology employed here allows us to investigate to what
extent, when quarks are confined within a hadron, the two
terms give equivalent results.

We emphasize that the rho electric dipole moment is studied
as a model problem.  The hadron electric dipole moment which
has been extensively studied experimentally and
theoretically is the neutron electric dipole moment.  The
relevant calculations are reviewed in \cite{HeMcKellarPakvasa89}.
We suggest that our results for the rho meson give an
indication of the effects that one may expect for other
hadron systems, including the neutron, when one has
available techniques for exploring the off mass shell
behaviour of the wavefunctions\footnote{A start was made on
that investigation by Costella\cite{costella}.}.

For a review of the topic of Dyson-Schwinger equations and
their application to hadron phenomenology see
\cite{RobertsWilliams94}.  For examples of the use of the
Dyson-Schwinger and Bethe-Salpeter equations in the
calculation of hadronic properties see references within
\cite{HechtMcKellar98,RobertsWilliams94}.  

As in our previous paper, use will be made of
forms for the quark propagator which ensure confinement by
having no Lehmann representation, {\em i.e.} the quarks
cannot go on mass-shell.  Three different phenomenological
forms for the rho amplitude will again be used.  In that
paper we used a dressed quark-photon vertex to ensure that
both the Ward-Takahashi Identity and multiplicative
renormalizability are preserved.  In our present work we
concentrate on the contribution of the $P$ and $T$
violating vertex function of Eq.~(\ref{new}), which we do
not dress.

\section{The ingredients of the calculation}
y\subsection{The $\rho\rho\gamma$ Vertex}
The form for the $\rho\rho\gamma$ vertex is given by
\cite{HechtMcKellar98,SalamDelbourgo64}
\begin{eqnarray}
V_{\alpha\mu\beta} & = & d_\alpha^{\ \nu}(p)\left[ P_\mu
      (-g_{\nu\sigma}{\cal E}(q^2)+q_\nu q_\sigma{\cal
Q}(q^2))\right.
      \nonumber \\
                              &   & +(g_{\mu\nu}q_\sigma
-g_{\mu\sigma}q_\nu )
      {\cal M}(q^2)
      +\varepsilon_{\nu\sigma\mu\rho}q^\rho
      \left.{\cal D}(q^2)\right] d_\beta^{\
\sigma}(p^\prime ), \end{eqnarray}
with $p$ the initial
momentum of the rho, $p^\prime$ the final momentum and $q$
the momentum of the photon.  The $d_{\alpha\beta}$ are the
rho spin projection operators
\begin{equation}
d_{\alpha\beta}(k) = g_{\alpha\beta}-\frac{k_\alpha
k_\beta}{k^2}.  \end{equation}
The form factors ${\cal
E}$, ${\cal Q}$, ${\cal M}$ and ${\cal D}$  are interpreted
 in the
limit as $q^2\rightarrow 0$ through $e{\cal E}(0) = e$, the
charge of the rho, $e{\cal M}(0)/(2m) = \mu$,
the magnetic moment, ${\cal Q}(0)$ is related linearly to
    $Q$, the   quadrupole moment,
and $e{\cal D}(0)/(2m) = d$, the electric dipole moment of the
rho.\footnote{We use $m$ without subscripts for the mass of the
$\rho^+$.  All other masses in this paper are identified with
the appropriate particle by subscripts.}  The
integral for the rho EDM is too difficult to perform
analytically.  This means that to isolate the relevant
terms they will  be projected out of the integral
before it is integrated numerically.  Using a projection
operator given by
\begin{equation}
P^{\alpha\mu\beta} =
-\frac{1}{2q^2}d^{\alpha\nu^\prime}(p)\varepsilon^{
        \mu\ \rho}_{\ \nu^\prime\
\sigma^\prime}d^{\beta\sigma^\prime}         (p^\prime),
\end{equation}
the following result holds
\begin{equation}
P^{\alpha\mu\beta}V_{\alpha\mu\beta}\stackrel{q\rightarrow
0}         {\longrightarrow}{\cal D}(0).
 \end{equation}

\subsection{Quark Contribution to the Electric Dipole Moment}
All calculations below are carried out in Euclidean space
with co-ordinates $x_1,x_2,x_3,x_4$, and metric and gamma
matrices given by $g_{\mu\nu} = \delta_{\mu\nu}$,
$\gamma_\mu = \gamma_\mu^\dagger$ and
$\{\gamma_\mu,\gamma_\nu\} = 2\delta_{\mu\nu}$. 
Treating the $u$ and $d$-%
quarks as identical except for their charge, the
impulse approximation to the quark contribution to the
electric moment is given by 
\begin{eqnarray}
\tilde{I}_{\alpha\mu\beta} & = & (-
1)\int^{}_{}\frac{d^4k}{(2\pi)^4}tr_{CFD}[
     \bar{\Gamma}^\rho_\beta (k;p-q)S(k_{-
+})\imath\tilde{\Gamma}_{\mu,q} (k_{++},k_{-+})
     \nonumber \\
                   &   &\times S(k_{++})\Gamma^\rho_\alpha
(k+q/2;p) S(k_{+-})].
\end{eqnarray}
where $k_{\alpha\beta} =
k+\frac{\alpha q}{2}+\frac{\beta p}{2}$,
$\Gamma^\rho_\alpha$ refers to the rho meson amplitude and
$\bar{\Gamma}^\rho_\alpha(k;p) =
C^\dagger\Gamma^\rho_\alpha(-k;p)C$ with $C$ the charge
conjugation operator, $\gamma_2\gamma_4$.  $\tilde{\Gamma}_{\mu,q}$ is
the $P$ and $T$ violating quark-photon vertex and $S(p)$ is the dressed
quark propagator for a quark of momentum $p$, all of which
will be discussed below.  The $tr_{CFD}$ operation is a
trace over colour, flavour and Dirac indices.  A tilde has
been placed on $I_{\alpha\mu\beta}$ to emphasize that this integral gives
only the $P$ and $T$ violating rho-photon vertex.

\subsection{Quark Propagators}
The general form for the solution to the quark propagator
Dyson-Schwinger equation \cite{RobertsWilliams94} is
\begin{eqnarray}
S(p) & = & -\imath \slsh
p\sigma_V(p^2)+\sigma_S(p^2) \nonumber \\
     & = & \left(\imath \slsh p A(p^2)+B(p^2)\right)^{-1}.
\end{eqnarray}
A model form for the propagator is given by
\cite{Roberts95,Roberts96}
\begin{eqnarray} \label{Ss-Sv}
\bar{\sigma}_S (x) & = & C_{\bar{m}}e^{-
2x}+\left(\frac{1-e^{-b_1x}}{b_1x}
     \right)\left(\frac{1-e^{-
b_3x}}{b_3x}\right)\left(b_0+b_2\left(
     \frac{1-e^{-\Lambda x}}{\Lambda x}\right)\right)
\nonumber \\
                   &   & +\frac{\bar{m}}{x+\bar{m}^2}
     \left(1-e^{-2(x+\bar{m}^2)}\right), \nonumber \\
\bar{\sigma}_V(x)  & = & \frac{2(x+\bar{m}^2)-1+e^{-
2(x+\bar{m}^2)}}
     {2(x+\bar{m}^2)^2}-\bar{m}C_{\bar{m}}e^{-2x},
\end{eqnarray}
where $x=p^2/2D,$ $\bar{\sigma}_V(x) =
(2D)\sigma_V(p^2),$ $\bar{\sigma}_S =
\sqrt{2D}\sigma_S(p^2)$ and $\bar{m} = m/\sqrt{2D},$ $D$
is a mass scale.  ($\Lambda = 10^{-4}$ is chosen to
decouple the small and large $\mbox{spacelike } p^2$
behaviour in Eq.~(\ref{Ss-Sv}); {\it i.e.} to allow $b_0$
to govern the ultraviolet behaviour and $b_2$ the
infrared.)  The parameters $C_{\bar{m}}$, $\bar{m}$,
$b_0,\ldots,b_3$ are \cite{BurdenRobertsThomson96},
$C_{\bar{m}\neq 0} = 0.0$, $C_{\bar{m} = 0} = 0.121$,
$\bar{m} = 0.00897$, $b_0 = 0.131$, $b_1 = 2.90$, $b_2 =
0.603$ and $b_3 = 0.185$, with the mass scale
$D=0.160\;{\rm GeV}^2$ chosen to give the correct value
for $f_\pi$. Note that with these values of the parameters
the mean light quark mass is $5.0$MeV.  
This form for the quark propagator is based
upon studies of the Dyson-Schwinger equation for $S(p)$
using a gluon propagator with an infrared singularity,
\begin{equation}
g^2D_{\mu\nu}(k) \equiv
\left(\delta_{\mu\nu}-\frac{k_\mu k_\nu}{k^2}\right)
  8\pi^4D\delta^4(k),
\end{equation}
and a dressed quark-gluon vertex \cite{Roberts96,BurdenRobertsThomson96}. A
sufficient condition for the lack of free quark production
thresholds is the absence of timelike poles in the
propagator.  The model quark propagator given above is an
entire function (except at timelike $p^2 = - \infty$) and so
does not have a Lehmann representation.  This means it can
be interpreted as describing a confined particle and it
ensures the lack of the unphysical singularities
corresponding to free quarks in $\tilde{I}_{\alpha\mu\beta}$.

The electric dipole moment is also calculated using a form
for the quark propagator developed by Mitchell and Tandy
\cite{MitchellTandy97}, to investigate $\rho$-$\omega$
mixing.  This propagator is given by
\begin{eqnarray}
\label{Ss-mitchell} \bar{\sigma}_S(x) & = & C_{\bar{m}}e^{-
2x}+\frac{\bar{m}}{x}\left( 1-e^{-2x} \right) ,
\\
 \label{Sv-mitchell}
\bar{\sigma}_V(x) & = & \frac{e^{-2x}-(1-2x)}{2x^2}-
\bar{m}C_{\bar{m}}e^{-2x}.
 \end{eqnarray}
To fix the
parameters $\lambda = \sqrt{2D}$ and $C_{\bar{m}}$, a fit
to $\langle \bar{q}q \rangle$, $f_\pi$, $r_\pi$ and the
$\pi$-$\pi$ scattering lengths was done.  With
$\frac{1}{2}(m_u+m_d) = 16$ MeV a best fit was obtained
for $\lambda = 0.889$ GeV and $C_{\bar{m}} = 0.581$
\cite{MitchellTandy97}.  This form has a deficiency which
can be seen in its failure to correctly model the
behaviour of $\sigma_S$ away from $x = 0$, in the massless
limit with dynamically broken chiral symmetry
\cite{Roberts96}.  The large value for the mass of the
quark used is related to this deficiency of the propagator
\cite{Mitchell97}.

\subsection{Rho Meson Amplitude}
The dominant Bethe-Salpeter amplitude for the rho meson is
given by \cite{PraschifkaCahillRoberts89,JainMunczek93},
\begin{equation}
\Gamma^{l}_{\rho\;\mu} (k,p) = \imath
\left(\gamma_\mu -\frac{\slsh p p_\mu}
     {p^2}\right)\tau^l\frac{\Gamma_\rho (k,p)}{N_\rho},
\end{equation}
where $k$ is the relative momentum of the
quark and anti-quark, $p$ is the momentum of the rho meson
and $l$ and $\mu$ are flavour and Dirac indices
respectively.  This form ignores other allowable Dirac
structure in the vector meson Bethe-Salpeter amplitude and
so introduces errors of the order of 10\%
\cite{FrankRoberts96}.  Using the quark propagator defined
in Eq.~(\ref{Ss-Sv}) and a Ball-Chiu quark-photon vertex
(see later), Chappell uses the following approximate form
for $\Gamma_\rho (k,p)$ \cite{Chappell96}
\begin{equation}
\Gamma_\rho = e^{-
k^2/a_1^2}+\frac{a_2}{1+\frac{k^2}{\alpha a_1}},
\end{equation}
with $a_1 = 0.38845$GeV, $a_2 = 0.01478,
\alpha = 2$GeV. 
 The values for the parameters were found by
fitting to the experimental values of $f_\rho$ and
$g_{\rho\pi\pi} $ \cite{Chappell96}.

Pichowsky and Lee,
also using the quark propagator defined in Eq.~(\ref{Ss-Sv})
and a quark-photon vertex of the Ball-Chiu type, use an
identical form for $\Gamma_\rho$, given by
\cite{PichowskyLee97}
\begin{equation}
\Gamma_\rho (k,p) =
e^{-k^2/a_V^2}+\frac{c_V}{1+k^2/b_V^2},
\end{equation}
where $a_V = 0.400$GeV, $b_V = 0.008$GeV and $c_V = 125.0$.
 Note that the coefficient of the rational term in the
vertex function, and the scale factor, differ dramatically
in the two attempts to fit this form of vertex function,

Mitchell and Tandy, with $S(p)$ defined by
Eqs.~(\ref{Ss-mitchell} and \ref{Sv-mitchell}), use a form for the
amplitude $\Gamma_\rho$ given by
\begin{equation}
\Gamma_\rho (k,p) = e^{-k^2/a^2},
\end{equation}
with $a =
0.194$ GeV.  The momentum scale in the exponential has been
reduced by a factor of $2$ and 
the rational term in the vertex function has
been omitted.

The normalisation for the rho amplitude is
fixed by \cite{PichowskyLee97}
\begin{eqnarray}
p_\mu
\left(\delta_{\alpha\beta}-\frac{p_\alpha
p_\beta}{p^2}\right) & = &         N_c{\rm
tr}_D\int_{}^{}\frac{d^4k}{(2\pi )^4}
        \frac{\partial S(k_+)}{\partial
p_\mu}\Gamma_{\rho\alpha}(k,p)S(k_-)
        \Gamma_{\rho\beta}(k,p) \\ \nonumber
 &  & + N_c{\rm tr}_D\int_{}^{}\frac{d^4k}{(2\pi )^4}
        S(k_+)\Gamma_{\rho\alpha}(k,p)
        \frac{\partial S(k_-)}{\partial
p_\mu}\Gamma_{\rho\beta}(k,p),
\end{eqnarray}
where
$k_{\alpha} = k+\frac{\alpha p}{2}$.  This condition,
with the fact that the quark-photon vertex
obeys the Ward Identity, ensures that ${\cal E}(q^2
= 0) = 1$, {\it i.e.} that the rho has unit charge
\cite{Roberts96}.

\subsection{The Quark-Photon Vertex}
The quark-photon vertex also satisfies its own Dyson-
Schwinger equation, but solving this integral equation is
difficult.  Despite this, a realistic ansatz for the
vertex function has been developed
\cite{BurdenRobertsWilliams93,DongMunczekRoberts94,BallChiu80,CurtisPennington90}.

 The
quark-photon vertex ansatz thus obtained is given by
\begin{equation}
\Gamma_\mu^{{\rm BC+CP}}(p,q) = Q_q \left[
\Gamma_\mu^{{\rm BC}}(p,q) +
     \Gamma_\mu^{{\rm CP}}(p,q)\right].
 \end{equation}
The
Ball-Chiu vertex, $\Gamma_\mu^{{\rm BC}}$ has the form
\cite{BallChiu80}
\begin{eqnarray}
\Gamma_\mu^{{\rm
BC}}(p,q) & = & \frac{A(p^2)+A(q^2)}{2}\gamma_\mu +
     \frac{(p+q)_\mu}{p^2-q^2}\left[\frac{1}{2}(A(p^2)-
A(q^2))(\slsh p +
     \slsh q)\right.  \nonumber \\
                      &   & -\left.\frac{}{}\imath
(B(p^2)-B(q^2))\right].
\end{eqnarray}
This vertex ansatz
is completely described by the dressed quark propagator
and satisfies both the Ward-Takahashi and Ward Identities,
is free of kinematic singularities as $q^2\rightarrow
p^2$, transforms correctly under appropriate
transformations and reduces to the perturbative result in
the appropriate limit.

To
ensure multiplicative renormalizability Curtis and
Pennington added a transverse piece to the Ball-Chiu
vertex \cite{CurtisPennington90}.  This term has the form
\begin{equation}
\Gamma_\mu^{{\rm CP}}(p,q) =
\left(\frac{-\imath\gamma_\mu (p^2-q^2)-(p+q)_\mu
     (\slsh p-\slsh q)}{2d(p,q)}\right)\left(A(p^2)-
A(q^2)\right),
\end{equation}
with
\begin{equation}
d(p,q)
= \frac{1}{p^2+q^2}\left[ (p^2-q^2)^2+\left(
M^2(p^2)+M^2(q^2)\right)^2
     \right],
\end{equation}
\begin{eqnarray}
M(p^2) =
\frac{B(p^2)}{A(p^2)}.  \nonumber
\end{eqnarray}

We introduce the quark electric dipole moment through the
vertex function $\tilde{\Gamma}^\mu$ of Eq.~(\ref{new}). We
have not attempted to ``dress'' this vertex by including
gluon loop corrections.

To have an explicit example in mind
we note that in models where the quark electric dipole moment
is generated by Higgs exchange\cite{lee,weinberg}, the
structure of the quark electric dipole moment is that of
Eq.~(\ref{new}). With the charged Higgs-quark interaction Lagrangian
given by
\begin{equation}
{\cal L}_{\mbox{int}}
 =   2^{3/4}G_F^{1/2}
\bar U \left[V_{KM}M_D \alpha
H^+ R  + M_UV_{KM}
\beta H^+ L\right]D
+ H.C.\;,
\end{equation}
the electric dipole moments of the up and down quarks are
dominated by the bottom and the top quark loops
respectively.  With a charged Higgs mass of the order of the
top mass, the electric dipole moment of the down quark is
significantly larger than that of the up quark, and the $P$
and $T$ violating $d-\gamma$ vertex is given by
Eq.~(\ref{new}), with
\begin{eqnarray}
d_{1,d} & = & K \left(F_1(m_t^2/M_H^2) +
F_2(m_t^2/M_H^2)\right) \label{d1} \\
d_{2,d} & = & - K F_1(m_t^2/M_H^2) \\
K & = & \frac{G_F
m_d}{48\pi^2}|V_{dq}|^2
\Im\left(\alpha^{*}\beta\right) \\
F_1  & = &  4\frac{x}{1-x}\left(1 + \frac{x}{1-x}\ln
x \right) \\
F_2 & =  & \frac{x}{\left(1-x\right)^2}\left(5
x -3 +\frac{2(3x - 2)}{1-x}\ln x\right), \label{f2}
\end{eqnarray}
Note that, in the on mass shell approximation, the $F_1$
terms cancel, leaving only the $F_2$ terms.

\subsection{Gauge Invariance, Ward Identities, Renormalization and 
Off-Shell behaviour}

We have already noted the the amplitude of (\ref{new}) is not 
manifestly gauge invariant, since
\begin{equation}
\partial_{\mu}   \tilde{\Gamma}^\mu_q   =  d_{1q} (p_{i}^{2} - 
p_{f}^{2}) \gamma_{5}
    \label{eq:div}
\end{equation}
which vanishes only when $p_{i}^{2} = p_{f}^{2}$, and thus vanishes on 
mass shell.   However, because $\tilde{\Gamma}^\mu_q$ is a Green's 
function, not an element of the $S$-matrix, it need only satisfy the 
Ward-Takahashi identity\cite{IZ},
\begin{equation}
\tilde{\Gamma}_\mu (p,p) = -\frac{\partial\Sigma_5(p)}{\partial 
p_\mu}, \label{eq:ward}
\end{equation}
or
\begin{equation}
    q_{\mu}\tilde{\Gamma}^\mu (p_{i},p_{f}) = 
    -\left(\Sigma_{5}(p_{i}) - \Sigma_{5}(p_{f}) \right).
    \label{eq:taka}
\end{equation}
The $CP$-violating term, $\Sigma_5(p)$, in the 
quark self energy,  required to satisfy the Ward Identity, will be 
generated by the same process that generates the $T$ violating 
vertex, for example by
Higgs loops.  In general 
\begin{equation}
    \Sigma_{5}(p) = m_{5}(p) \gamma_{5},
    \label{eq:tvse}
\end{equation} and 
$\Sigma_5(p)$ is divergent, coming from a one loop diagram.  There are 
however no terms of this structure in the unrenormalised lagrangian,
so we have 
no parameter to adjust in the renormalisation process.  In $QED$ this 
is not a problem, as one can demand that the equivalent term vanish 
on mass shell\cite{FKW}.  In the present case quarks do not have a mass
shell in the confined phase of $QCD$, and the best we can do 
is to renormalize $\Sigma_{5}$ such that it
vanishes at some renormalization point, $\mu$.
That is, we have, after  renormalisation,
\begin{equation}
\Sigma_5(p) = -d_1(p^2-\mu^2)\gamma_5,
\end{equation}
guaranteeing the Ward-Takahashi identities\footnote{At least when we 
can regard $d_{1}$ as independent of $p_{i}$ and
$p_{f}$}. For non-confined particles the choice
$\mu = m$, where $m$ is the mass of the particle
concerned, recovers the results of 
\cite{FKW}. 

\begin{figure}[h]
\psfig{file=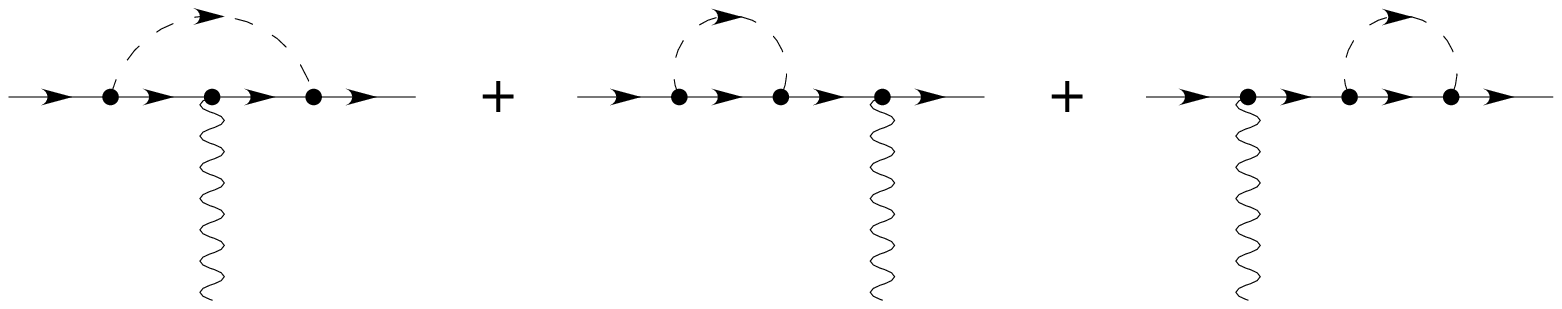,width=15cm,height=3cm}
\caption{Lowest order Higgs loop corrections to the quark-photon vertex}
\label{higgs-corrections}
\end{figure}

To lowest order then, the corrections to the quark-photon vertex due to
Higgs loops are given by the sum of diagrams given in 
Fig.~(\ref{higgs-corrections}). If $\mu$ is set equal to $m_q$ then the
``non-standard'' quark-photon vertex term in Eq.~(\ref{new}) is cancelled,
with the result that
\begin{equation}
\tilde{\Gamma}_q^\mu = (d_2+d_1)\imath\sigma^{\mu\nu}q_\nu\gamma_5,
\end{equation}
just as if we had used the on-shell identity of 
Eq.~(\ref{gordon}).
If, on the other hand, one renormalizes $\Sigma_5$ at 
some other point $\mu\neq m_q$,
then there is no such cancelation but $\tilde{\Gamma}_q^\mu$ has a $\mu^2$
dependence. This $\mu^2$ dependence is a result of the absence of an 
adequate theory of perturbations about the confined phase, indeed of 
an absence of an adequate theory of the confined phase. This being the case,
to make progress we set set $\mu^2 = \langle p_{av}
\rangle^2$, with the result that the last two diagrams of 
Fig.~(\ref{higgs-corrections}) give negligible contributions and so
\begin{equation}
\tilde{\Gamma}^\mu_q = d_1(p_i+p_f)^\mu\gamma_5+d_2\imath\sigma^{\mu\nu}
	q_\nu\gamma_5,
\end{equation}
just as Eq.(\ref{new}).

It may be thought that the lack of gauge invariance of the $CP$ 
violating vertex of $(\ref{new})$ would lead to a lack of gauge 
invariance for the electric dipole $\rho\rho\gamma$ coupling.  
Explicit calculation shows this not to be the case --- the 
resulting electric dipole $\rho\rho\gamma$ coupling has  only the gauge 
invariant form $\varepsilon_{\nu\sigma\mu\rho}q^\rho$.  While a 
calculation of the neutron EDM using the 
$T$ and $P$ violating vertex of Eq.~(\ref{new}) in a three quark 
model of the nucleon is at present impractical, if one uses 
it in a quark-diquark 
model of the nucleon,  the resulting $NN\gamma$,  $T$ and $P$ violating, 
vertex is similarly gauge invariant.

\section{The Dipole Moment}
The dipole moment of the rho can now be calculated.  The
colour, flavour and Dirac traces are performed for the
integral $\tilde{I}_{\alpha\mu\beta}$, see appendix A, and the
dipole term projected out using the projection operator
$P^{\alpha\mu\beta}$.  The integral is then performed
numerically using Gaussian quadrature methods to obtain
the results given in Table~\ref{table1}.

For comparison, we also use perturbation theory to
calculate the electric dipole moment of a rho meson made
of free, undressed quarks.  To do this for the $(p_i +
p_f)^\mu\gamma_5$ term\footnote{the equivalent calculation
for the $\imath \sigma^{\mu\nu}q_\nu\gamma_5$ contribution
was performed in our earlier paper\cite{HechtMcKellar98},
with the result that the rho electric dipole moment is
$d_\rho = (m_ud_u + m_d d_d)/(2m)$  in this
perturbative limit.}, in the integral
$\tilde{I}_{\alpha\mu\beta}$, the following replacements
are made,
\begin{eqnarray}
S(p) & \rightarrow &
1/(\imath\slsh p+m) \\
 \Gamma^\rho(k,p) & \rightarrow & 1 \\
\tilde{\Gamma}_\mu(\kpp,\kmp) & \rightarrow &
(\kpp+\kmp)_\mu\gamma_5 D_{1,q}Q_q.
 \end{eqnarray}
  Using the standard dimensional regularization scheme
along with the Feynman parameterization technique
\cite{MandlShaw}, the integral can be shown to remain
finite.  However, as was shown in our earlier
paper~\cite{HechtMcKellar98}, the corresponding contribution to the
electric charge is infinite.  Thus calculating the
contribution to the rho dipole moment in perturbation
theory via the ratio ${\cal D}(0)/{\cal E}(0)$ yields
\begin{equation} d_{\rho;{\rm pert}} = 0.  \end{equation}

In \cite{HechtMcKellar98} the rho EDM was
also calculated in the bag model.  As it has been noted
previously that $\bar{\psi}(p_i+p_f)_\mu\gamma_5\psi =
\bar{\psi}\imath\sigma_{\mu\nu}q^\nu\gamma_5\psi$ for
quarks on the mass-shell, it follows that the non-standard
form for the quark EDM will yield the same bag model
result as the standard term, since the quarks satisfy the
equation of motion inside the bag.

In the non-relativistic limit, the result is simply
\begin{equation}
d_\rho^{(NR)} = \left\{\left(d_{1,u} +
d_{2,u}\right) + \left(d_{1,d} + d_{2,d}\right)
\right\}.
\end{equation}

The results
may be expressed in the form
\begin{equation}
d_\rho = \left( d_{1,u} + 
d_{1,d}\right) A + \left( d_{2,u} +
 d_{2,d}\right) B
\end{equation}

For our various choices of input, A and B are given in Table
\ref{table1}.

\section{Discussion}
Using a model form for the quark propagator obtained from the
Dyson- Schwinger equations, two different phenomenological
rho-meson amplitudes fitted to $f_\rho$ and
$g_{\rho\pi\pi}$ and one used to describe $\rho$-$\omega$
mixing, and a quark-photon vertex which incorporated the
$P$ and $T$ violating quark-photon vertex function of
Eq.~(\ref{new}), we have calculated the electric dipole moment
of the rho.

For each of the coefficients $A$ and $B$, 
all three non-perturbative models agree qualitatively with
the Bethe-Salpeter amplitude of \cite{PichowskyLee97}
yielding the largest value for $d_\rho$ and that of
\cite{MitchellTandy97} the smallest.  The results for the
amplitudes of Chappell, and Pichowsky and Lee, are very
close, which is indicative of the similarities of the two
approaches.

For the same non-perturbative model, a comparison of the
coefficients $A$ and $B$ shows that the contribution of the
$(p_i+p_f)_\mu \gamma_5$ term  is rather less than that of
the
``standard'' $\imath\sigma_{\mu\nu}q^\nu\gamma_5$
 term.
The $(p_i+p_f)_\mu\gamma_5$ contribution was
$\sim\!\!67\%$ of the $\imath\sigma_{\mu\nu}q^\nu\gamma_5$
result for the amplitude developed by Mitchell and Tandy,
$\sim\!\!83\%$ for that of Chappell and $\sim\!\!81\%$ for
the amplitude of Pichowsky and Lee, with the
$(p_i+p_f)_\mu\gamma_5$ and
$\imath\sigma_{\mu\nu}q^\nu\gamma_5$ terms weighted
equally. 

In the Bag Model, and in the non-relativistic limit, these
two terms contribute equally.  In the perturbative, free
limit, on the other hand the
$\imath\sigma_{\mu\nu}q^\nu\gamma_5$ term gives a finite
(but small) contribution, whereas the $(p_i+p_f)_\mu
\gamma_5$ term gives a vanishing contribution.  In a sense
the non-perturbative models interpolate between these
limits.

To be explicit, consider the Higgs model introduced above.
 We take the Higgs mass to be $100$GeV, and the top mass
to be $170$GeV, and note that, in this model, 
 $d_d \gg d_u$.  Eqs.~(\ref{d1})-(\ref{f2})  show that the
$(p_i+p_f)^\mu\gamma_5$ term has a coefficient more than twice as
large as that of the standard term, {\em i.e.} $d_{1,d}/d_{2,d} = 2.34$.
  The resulting rho
electric dipole moment, expressed as a fraction of the
non-relativistic value is given in Table \ref{table2}.

In this case the models of Chappell, and Pichowsky and
Lee, give similar results which are about two  thirds of the
non-relativistic result.  The Mitchell and Tandy result is
significantly smaller.  The good news is that the
standard non-relativistic result is good to a factor of $2$
to $3$, and that while one is concerned with approximate
estimates it is a reasonable guide to the overall effect.

Though this calculation was carried out for the rho meson
it leads to some interesting speculation about the 
calculation of the quark contribution to the neutron
electric dipole moment, which is of interest in
the study of CP-violation.  The use of the full off mass
shell $P$ and $T$ violating interaction of Eq.~(\ref{new})
could significantly alter both the standard model
and the many non-standard model results for the neutron EDM.
While one can expect the order of magnitude of the existing
calculations to be correct, at a precision level the off-mass
shell effects considered here will become important. 
As some of
the non-standard model calculations yield values that are
close to the current experimental limits
\cite{HeMcKellarPakvasa89}, it will become important to
attempt precision calculations in the future.  Of course,
when such calculations are attempted, many additional
sources of a neutron electric dipole moment will need to be
considered.  In the Higgs model, these include
\begin{enumerate}
\item charged Higgs contributions to
\begin{enumerate}
\item the quark colour-electric dipole moment
\item the gluon colour-electric dipole moment
\item other $P$ and $T$ odd quark-quark interactions
\end{enumerate}
\item   neutral Higgs contributions to the above
\item hadronic effects
\end{enumerate}
Considerations similar to those discussed here will enter
the calculation of the quark colour-electric dipole moment
effects, and possibly also the other additional effects.

It is well known that the fact that the quarks in hadrons
are both relativistic and bound, and thus are significantly
off their mass shell, has significant effects on the quark
contribution to the magnetic moments of
hadrons\cite{HechtMcKellar98,TupperMcKellarWarner88}.
The study of the rho meson presented 
here shows that similar effects occur in the
calculation of the electric dipole moments of hadrons.  In
particular, it is necessary to differentiate between the two
forms of the $P$ and $T$ odd quark-photon interaction in
these calculations, and not to treat them as being
equivalent as they are on the quark mass shell.  The next
generation of calculations of the electric dipole moment of
the neutron should take such effects into account.

\appendix

\section{Integral for the Electric Dipole Moment}
Taking the colour and flavour traces, the integral becomes
\begin{eqnarray}
\tilde{I}_{\alpha\mu\beta} & = & \frac{2\imath N_c}{N_\rho^2}\int^{}_{}
     \frac{d^4k}{(2\pi )^4}\Gamma_\rho (k)\Gamma_\rho (k+q/2)tr_D\left[
     \frac{}{}\right.  \nonumber \\
     &   & +\gb\skmp\gfv\skpp\ga\ktldm T_1
	+\gb\skmp\gfv\ga\skpm\ktldm T_2
	\nonumber \\
	&   & +\gb\skmp\gamma_5\sigma_{\mu\rho}q^\rho
    \skpp\ga T_3+
    \gb\skmp\gamma_5\sigma_{\mu\rho}q^\rho\ga\skpm T_4
\nonumber \\
     &   & +\gb\gfv\skpp\ga\skpm\ktldm T_5\nonumber \\
&   & +\gb\gamma_5\sigma_{\mu\rho}q^\rho\skpp
    \ga\skpm T_6+
    \gb\gamma_5\sigma_{\mu\rho}q^\rho\ga T_7
	\left.\frac{}{}\right],
\end{eqnarray}
where $\ktldm = (\kpp+\kmp)_\mu$, only $P$ and $T$ odd terms
have been retained, and terms which will not survive the
Dirac trace have been omitted.
  The $T_{1-7}$ are defined by
\begin{eqnarray}
T_1 & = & -V_1\svkmp\svkpp\sskpm, \nonumber \\
T_2 & = & -V_1\svkmp\sskpp\svkpm, \nonumber \\
T_3 & = & -V_2\svkmp\svkpp\sskpm, \nonumber \\
T_4 & = & -V_2\svkmp\sskpp\svkpm, \nonumber \\
T_5 & = & -V_1\sskmp\svkpp\svkpm, \nonumber \\
T_6 & = & -V_2\sskmp\svkpp\svkpm, \nonumber \\
T_7 & = & V_2\sskmp\sskpp\sskpm,
\end{eqnarray}
with,
\begin{eqnarray}
V_1 & = & -\left(d_{1,u} + d_{1,d}\right),
\nonumber \\
V_2 & = & -\imath \left(d_{2,u} +
d_{2,d}\right).
\end{eqnarray}

\newpage

\begin{table}[h]
\caption{Expansion coefficients $A$ and $B$ for the
Electric Dipole Moment $d_\rho$
}
\label{table1}
\begin{center}
\begin{tabular}{|c|c|c|}
Model & A & B\\
\hline
Chappell & 0.620 & 0.743 \\
Pichowsky and Lee & 0.627 & 0.779\\
Mitchell \& Tandy & 0.418 & 0.627\\
Perturbative & 0.000 & 0.010 \\
Bag Model & 0.828 & 0.828\\
Non-Relativistic & 1.000 & 1.000\\
\end{tabular}
\end{center}
\end{table}

\newpage

\begin{table}[h]
\caption{Electric Dipole Moment $d_\rho$ in a Higgs model,
as a ratio to the Non-Relativistic result}
\label{table2}
\begin{center}
\begin{tabular}{|c|c|}
Model & Ratio \\
\hline
Chappell & 0.657 \\
Pichowsky and Lee & 0.672\\
Mitchell \& Tandy & 0.465\\
Perturbative & 0.003\\
Bag Model & 0.828\\
Non-Relativistic & 1.00\\
\end{tabular}
\end{center}
\end{table}

\newpage


\newpage
\begin{thebibliography}{99}

\bibitem{HechtMcKellar98}
M.B.  Hecht and B.H.J.  McKellar, Phys. Rev. C {\bf C57} 2638 
(1998).

\bibitem{HeMcKellarPakvasa89}
X-G.  He, B.H.J.  McKellar, and S.  Pakvasa,
Int. J. Mod. Phys. A
{\bf 4} 5011 (1989).

\bibitem{RobertsWilliams94}
C.D.  Roberts and A.G.  Williams, Prog. Part. Nucl. Phys {\bf 33} 24 (1994).

\bibitem{costella}
J. P. Costella, B.Sc.(Hons.) Thesis ``The Neutron Electric Dipole
Moment in a Relativistic Constituent Quark Model'', University of Melbourne
(1990).

\bibitem{SalamDelbourgo64}
A.  Salam and R.  Delbourgo, Phys. Rev. {\bf 135} 1398 (1964).

\bibitem{Roberts95}
C.D.  Roberts, in ``Chiral Dynamics: Theory and Experiment'', Bernstein, A.M.
and Holstein, B.R.  (Eds.), Lecture Notes in Physics, Vol.  452, p.  68
(Springer, Berlin 1995).

\bibitem{Roberts96}
C.D.  Roberts, Nucl. Phys A {\bf 605} 475 (1996).

\bibitem{BurdenRobertsThomson96}
C.J.  Burden, C.D.  Roberts and M.J.  Thomson, Phys. Lett. B {\bf 371}
163 (1996).

\bibitem{MitchellTandy97}
K.L.  Mitchell, P.C.  Tandy, Phys. Rev. C
{\bf 55} 1477 (1997).

\bibitem{Mitchell97}
K.L.  Mitchell, Private Communications (1997).

\bibitem{PraschifkaCahillRoberts89}
J.  Praschifka, R.T.  Cahill and C.D.  Roberts, Int. J. Mod. Phys. A
{\bf 4} 4929 (1989).

\bibitem{JainMunczek93}
P.  Jain and H.  Munczek, Phys. Rev. D {\bf 48} 5403 (1993).

\bibitem{FrankRoberts96}
M.R.  Frank and C.D.  Roberts, Phys. Rev. C {\bf 53} 390 (1996).

\bibitem{Chappell96}
I.  Chappell, Private Communications (1996).

\bibitem{PichowskyLee97}
M.A.  Pichowsky and T.-S.H.  Lee, Phys. Rev. D {\bf 56} 1644 (1997).

\bibitem{BurdenRobertsWilliams93}
C.J.  Burden, C.D.  Roberts and A.G.  Williams, Phys. Lett. B
{\bf 285} 347 (1992).

\bibitem{DongMunczekRoberts94}
Z.  Dong, H.J.  Munczek and C.D.  Roberts, Phys. Lett. B
{\bf 333} 536 (1994).

\bibitem{BallChiu80}
J.S.  Ball and T.W.  Chiu, Phys. Rev. D {\bf 22} 2542 (1980).

\bibitem{CurtisPennington90}
D.C.  Curtis and M.R.  Pennington, Phys. Rev. D {\bf 42} 4165 (1990).

\bibitem{lee} T. D. Lee, Phys. Rev. D {\bf 8},
1226 (1973); Phys. Rep. {\bf 96}, 143 (1974).

\bibitem{weinberg} S. Weinberg, Phys. Rev. Lett. {\bf
37}, 657 (1976).

\bibitem{IZ} C. Itzykson and J. Zuber, ``Quantum Field Theory'', McGraw-Hill,
New York, 1980.

\bibitem{FKW} G. Feinberg, P. Kabir, and S. Weinberg, Phys. Rev. Lett. 
{\bf 3}, 527 (1959). 

\bibitem{MandlShaw}
F.  Mandl and G.  Shaw, ``Quantum Field theory'', John Wiley \& Sons, New York 
1993.

\bibitem{TupperMcKellarWarner88}
N.E.  Tupper, B.H.J.  McKellar, and R.  Warner,
Australian Journal of Physics {\bf 41} 19 (1988).

\end{thebibliography}
\end{document}